# Excited state spectroscopy in carbon nanotube double quantum dots

*Sami Sapmaz\*, Carola Meyer[†], Piotr Beliczynski, Pablo Jarillo-Herrero, and Leo P. Kouwenhoven*

Kavli Institute of Nanoscience, Delft University of Technology, P.O. Box 5046, 2600 GA Delft, The Netherlands

sami@qt.tn.tudelft.nl

[†]new address: Research Centre Jülich, Institute of Solid State Research, Electronic Properties, 52425 Jülich, Germany

We report on low temperature measurements in a fully tunable carbon nanotube double quantum dot. A new fabrication technique has been used for the top-gates in order to avoid covering the whole nanotube with an oxide layer as in previous experiments. The top-gates allow us to form single dots, control the coupling between them and we observe four-fold shell filling. We perform inelastic transport spectroscopy via the excited states in the double quantum dot, a necessary step towards the implementation of new microwave-based experiments.

Electron spins in double quantum dots (DQDs) are one of the leading systems for fundamental studies of elementary solid-state qubits[1]. Recent progress has been based on DQDs in 2-dimensional electron gases in GaAs semiconductor heterostructures[2,3]. However, the presence of non-zero nuclear spins limits the decoherence time in such structures[3,4]. This drawback has stimulated the quest for novel materials that allow to fabricate DQDs with longer spin decoherence times. Among these, carbon nanotubes (CNTs) have properties that seem to make them an ideal material. Most of the natural carbon (98.93%) is $^{12}$C without nuclear spin. Thus, the nuclear field that leads to spin relaxation[5,6] will be small compared to III/V semiconductors or even zero, if pure $^{12}$C is used in the growth of the carbon nanotubes. Also the spin-orbit interaction, which limits the spin relaxation time in semiconductor heterostructures[6,7,8], is expected to be very small in CNTs because of the low atomic number of carbon.

In order to realize CNT-DQDs (Figure 1), it is highly desirable to be able to create tunable tunnel barriers at arbitrary locations in a CNT, and some elementary devices have already been demonstrated[9,10]. In order to make use of a CNT-DQD in quantum information processing, e.g. by determining spin and orbital relaxation times and by performing quantum operations, access to (spin) excited states is crucial. These excited states have not been observed previously in CNT DQDs. Here we demonstrate an improved fabrication scheme compared to earlier approaches. Thin top-gates are evaporated such that only a small portion of the CNT is covered with oxide. We show electronic transport through the ground and excited energy states of CNT-DQDs.

The single walled carbon nanotubes (SWCNTs) are grown using chemical vapor deposition[11] (CVD) at lithographically predefined positions on a degenerately p-doped Si substrate with a 250 nm thick thermal oxide using a Fe/Mo based catalyst. The nanotubes are located by atomic force microscopy with respect to predefined markers, such that contacts and gates can be designed for each tube individually. First, we fabricate the contacts by means of electron beam lithography on a double layer PMMA-based resist, followed by metal (Pd) evaporation and lift-off in acetone. We use Pd as contact material because it introduces little or no barrier at the nanotube-metal contact[12,13]. In a subsequent electron beam lithography step we 'write' the gate structures. For the top-gates we evaporate a 2 nm thin layer of Al and use the natural formation of thin insulating native oxide in an oxygen environment. We oxidized for ten minutes at 1 bar pure oxygen pressure to ensure a 2 nm thick oxide. These steps are then repeated once more, so that we end up with an oxide thickness of 4 nm. The evaporation is continued with a 35 nm thick Al layer and finalized by 15 nm of AuPd on top. The top-gates are about 30 nm wide, which is the lower limit for our fabrication process. The advantage of a narrow top-gate is that it controls the tunneling barrier on a local scale and only a small portion of the tube is covered with oxide. We think that this is an advantage for future devices, since an oxide always has charge traps and therefore provides a source of charge fluctuations that interfere with transport measurements.

Figure 1 shows an AFM image of a representative sample. The nanotube is divided into four segments by the three top-gates ($TG_L$, $TG_M$ and $TG_R$). The SWCNT segments between source-$TG_R$ and $TG_L$-drain are not forming a quantum dot due to the low source and drain contact resistances[13]. By applying voltages to the top-gates we can tune the barriers and create quantum dots. Each quantum dot is addressed individually by the side-gates $SG_L$ and $SG_R$. Room-temperature measurements show a source-drain resistance of 30 k$\Omega$ and a small variation of the resistance with changing back-gate voltage indicating a small band gap tube. The conductance versus the left side-gate voltage for different values of the central top-gate is shown in the supplementary information.

We can operate the sample in both the p-doped and n-doped regions. At low temperatures (300 mK) we observe the highest conductance when applying negative voltages to the top-gates and operate the device in the hole-transport regime. Figure 2 shows a differential conductance plot as a function of source-drain and back-gate voltage. The highly-doped silicon substrate is used as the back-gate with the intention to change the electrochemical potential of the nanotube uniformly. Note that the average conductance is between 2 and 3 times $e^2/h$ (the measured maximum is 3.14 $e^2/h$). The pattern in Figure 2 is due to quantum interference in the nanotube which acts as an electron waveguide in analogy with the optical Fabry-Perot cavity, as previously studied in nanotubes[13,14]. The bias voltage at the crossing point $V_C$ between adjacent left- and right-sloped dark lines (see white arrows) is found to depend on the length of the waveguide L, as $V_C = hv_F/2eL$ (Fermi velocity $v_F = 8.1 \ 10^5$ m/s [15]). The pattern in figure 2 is less regular than reported in previous studies[14]. A possible reason for this is the presence of narrow top-gates on the tube which can act as weak scatterers. We mainly find a value of $V_C = V_{SD} \sim 0.5\text{-}0.6$ meV, corresponding to a length of about 3 $\mu$m. This suggests that the electron scattering occurs primarily at the nanotube-Pd interfaces since the extracted length is much larger than the top-gate spacing. The overestimation of the length could be due to band bending. Close to the band gap, the dispersion dE/dk is smaller than in the linear dispersion relation of a metal tube and therefore the velocity decreases as well, and thus results in a smaller value for the crossing point $V_C$.

By applying positive voltages to the top-gates we form barriers in this p-type region. Figure 3a shows the typical characteristics of a clean nanotube single quantum dot. We form the dot by setting the voltages on $TG_L$, $TG_M$ to 4 V, and $TG_R$ to -4 V to keep this part open. The four-fold shell filling pattern in the Coulomb blockade diamonds expected for carbon nanotubes[16,17,18,19] is clearly visible (see inset for addition energies) together with the excitation lines (lines parallel to the diamond slope). Only 5 parameters are needed to characterize the nanotube low-energy electronic shell structure[20]. These are the charging energy $E_C$, the orbital level spacing $\Delta$, the sub-band mismatch $\delta$, the exchange energy $J$, and the excess Coulomb energy $dU$. The fact that the three smaller diamonds in Figure 3a are about the same size (see also inset) and quite smaller than the large diamond suggests that the values for $\delta$, $J$, and $dU$ are small compared to $\Delta$. We extract for $E_C \sim 4$ meV, and for $\Delta \sim 2$ meV. This is also consistent with the observed excitation lines. A clear excitation line is present at around 2 meV and additionally we see inelastic co-tunneling lines[21] at energies less than 200 μeV, which are most likely due to the small values of the parameters $\delta$, $J$, $dU$. We obtain the length of the nanotube quantum dot $L = hv_F/2\Delta = 740$ nm from the value of the level spacing (~2 meV). This is in reasonable agreement with the fabricated nanotube quantum dot size of 550 nm. The overestimation of its length may again be due to the smaller velocity close to the gap (see discussion above).

The characterization of the right dot is shown in Figure 3b. No clear shell-filling pattern is observed in this case. But closed diamonds and constant slopes of the diamond edges indicate that we measure a single dot. Excited states are also clearly visible. Assuming that $\delta$, $J$, and $dU$ are as small as in the left dot, we obtain a larger charging energy in the order of $E_C \sim 10$ meV [22]. From the excited states, we find the level splitting $\Delta \sim 3$ meV. This corresponds to a quantum dot length of 570 nm, again in good agreement with the designed size. We also observe negative differential conductance (NDC) which is seen as black lines parallel to the diamond slope. NDC has been observed in many types of quantum dots and can have different origins[23,24,25,26]. A possible explanation in our structure could be that a poorly coupled excited state becomes occupied, which then blocks the current through the ground state. This leads to a smaller current at larger bias and therefore to NDC.

Figure 4 shows the characteristic "honeycomb" structure of the current through a double quantum dot[27] in the strongly coupled regime. Here, the two dots are not completely separated but interact via tunnel coupling, thus forming the analogue of a molecule with covalent bonding. The co-tunneling lines of the hexagonal pattern are visible and exhibit the four-fold shell filling for the left dot, just as the diamonds did in the previous single dot measurement: a large hexagon is followed by three small ones in the vertical direction of the left side-gate. This pattern repeats for every electron number in the right dot with the same top-gate settings.

In Figure 5 we show the double dot in the weakly coupled regime, i.e. the inter-dot tunnel resistance is high and the capacitive coupling between the dots dominates the transport behavior. The measurements are done in a different gate-region than the previous measurements. A small voltage is applied at the center top-gate (200 mV) and the left and right top-gates are set to zero Volts in order to reach the weakly coupled regime. The values for the side-gates are adjusted because the top-gates do not only change the tunnel barriers but shift the chemical potential of the dot as well. The triple points of the expected hexagonal pattern are very well visible and, due to the large bias, develop into overlapping triangles[27]. Excited states are observed in every triple point. In the supporting information we show measurements on a different nanotube double dot in figure 3 to demonstrate reproducibility.

All capacitances[27] that characterize the double quantum dot are calculated from the size of the hexagons and triple points in Figure 5: The capacitive coupling between the dots is $C_{TGM} = 1$ aF, the total capacitances of the left and right dot are $C_L = 7.2$ aF and $C_R = 12.5$ aF, respectively, and the relative capacitances between each side-gate and its neighboring dot are $C_{SGL} = 3$ aF and $C_{SGR} = 1.5$ aF.

We obtain tunnel barriers even for zero or small top-gate voltages for certain gate regions. This indicates that the fabrication of the local top-gates alone induces a small barrier. A small co-tunneling current is visible in the upper right area of Figure 5 which shows up as lines linking the triple points. The fact that there is no co-tunneling current in the rest of the figure shows that the tunnel barriers change while changing the side-gate voltages and that there are cross capacitances between the gates.

A high resolution measurement of a pair of triple points (electron- and hole-cycle) is shown in Figure 6 for $V_{SD} = 4$ mV. At the baseline of the triangle the ground states of the two dots are aligned and shifted together from the Fermi-level of the drain (point **a** in Figure 6) to the Fermi-level of the source (point **b**). At the center of the baseline they lie exactly in the middle between source and drain. On a line from this point to the tip of the triangle (point **c**), the states of the right dot are shifted downwards to the Fermi-level of the source, while the states of the left dot shift upwards to the drain (a positive bias is applied at the source contact, while the drain contact is put to ground). Along this line, we see sharp excitations at 0.33, 1.24, 1.55, and 1.8 mV (see inset of Figure 6). These lines belong to different excited states of the left dot which are probed by states of the right dot. An area of non-resonant current spreads between 2 and 2.8 mV. In figure 2 of the supporting information we show data for an adjacent pair of triple points. The excited states for both pairs are consistent.

In the following we give a possible scenario for the resonant transport. Afterwards, we discuss the non-resonant current. On the right side of the triangles of both triple points there appears to be a region of strongly suppressed current. This feature could be explained by bad coupling of the ground state of the right dot to the source. As the levels of the dots move upwards with lowering the side-gate voltages, at point (*) the first hole-excited state[28] of the right dot enters the bias window at ∼ 650 μeV. The coupling of this level to the source contact is stronger, thus enhancing the current.

The lines parallel to the baseline of the triangle belong to resonant transport through hole-excited states of the left dot. Only the first of these excited states at ∼330 μeV is probed by the ground state of the right dot (point **e**). At the other lines, the excited state of the right dot at ∼650 μeV is aligned with the excited states of the left dot (see point **f** as an example). This could also explain the larger current through these lines, as the excited states are probably better coupled to the contacts. Taking this into account, the energy splitting of the second excited state (**) of the left dot to its ground state is 1.9 meV. This fits well with the value for the level splitting $\Delta$ we obtained from the single dot measurement. The next two lines split by 310 and 560 μeV with respect to the level (**) at $\Delta$. Like the first one at ~330 μeV (point **e**) they are comparable in size with the low energy splitting found in the single dot measurement of the left QD.

Non-resonant transport can occur if an electron looses energy due to spontaneous emission of an acoustic phonon[29]. However, we do not observe the expected decay of the current for one-dimensional acoustic phonons with detuning of the DQD states. The non-resonant current between 2 mV and 2.8 mV seems to have its origin rather in level broadening of excited states at higher energy.

Electron-phonon coupling in a molecule such as a CNT can show up as sharp resonance lines. These would be equidistant with an energy difference that depends on the diameter and length of the tube[30]. For a length of 2 μm, an energy difference of $\Delta E_{phonon}$ ~55 μeV is expected. If the size of the single QDs of ~ 500 nm would determine the energy of the phonons, one would expect $\Delta E_{phonon}$ ~ 440 μeV. None of these energy scales show up in the lines inside the triangle. Thus we conclude that the lines inside the triangle are due to resonant transport through electronic excitations.

In this letter we have demonstrated the full tunability of a carbon nanotube double quantum dot. We avoid covering the whole tube with an oxide layer by fabricating narrow $AlO_x/Al$ top gates, and by that disturbing the structure of the carbon nanotube as little as possible. In this way, we are able to show Fabry-Perot interference between the source/drain contacts in a SWCNT structure with top-gates. The typical four-fold shell filling of a small band gap tube and excited states are visible in the stability diagram of the left dot. This shows that we are able to fabricate tunable barriers for carbon nanotube quantum dots, which is essential for spin relaxation time measurements[31,32]. The shell filling of the single dot is also visible in the hexagon pattern of the double quantum dot. The excited states of both dots show up as resonant tunneling lines in the triple points. This paves the way for new microwave-based quantum information processing experiments with carbon nanotubes.

**Acknowledgment.** We thank Frank Koppens for discussions and C. Dekker for the use of CNT growth facilities. This work was supported by the Defense Advanced Research Projects Agency Quantum Information Science and Technology program, the Dutch Organization for Fundamental Research on Matter (FOM), the Netherlands Organization for Scientific Research (NWO), and the EU Research Training Network on spintronics.

**Supporting Information Available.** Additional measurements are presented in the supporting information. We show the conductance versus the side-gate voltage for different values of the central top-gate. Furthermore, a pair of triple points adjacent to the one from figure 6 is shown and its excited states are described. Agreement in behaviour with figure 6 is found. Finally, to demonstrate reproducibility, we show measurements on a different nanotube double dot, which is operated in the weak- and strong tunnel coupled regime.

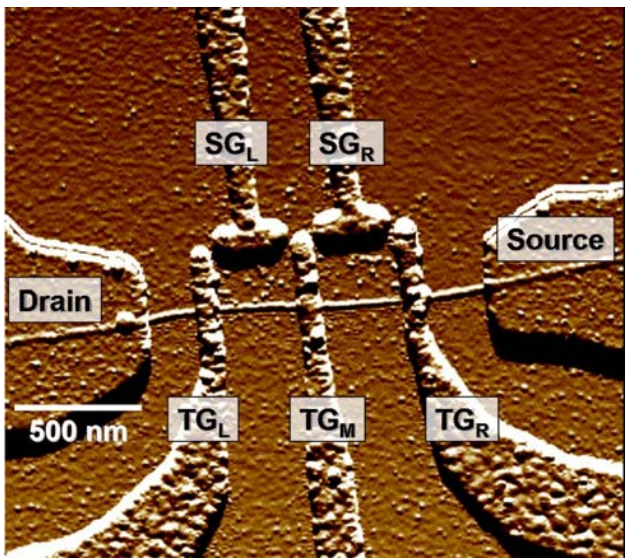

**Figure 1.** AFM picture of a CNT DQD device similar to the one used for the actual measurements. The actual device (not shown here) has a tube diameter of about 3nm and a total length of 1.9 μm between the Pd contacts, the left and right dots are both 550 nm, source-$TG_R$ segment is 560 nm long and the remaining $TG_L$-drain distance is 220 nm. Room-temperature measurements show a source-drain resistance of 30 k$\Omega$ and a small variation of the resistance with changing back-gate voltage. For the three top- and the two side-gates, a layer structure of $Al_2O_3$/Al (4 nm / 35 nm) has been used. All structures in the image appear wider due to AFM tip convolution.

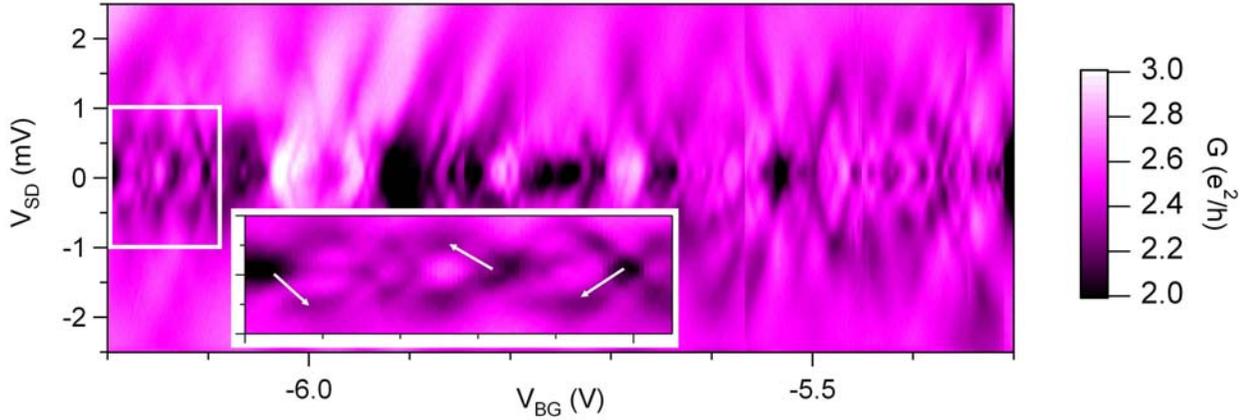

**Figure 2.** Differential conductance measured with a lock-in amplifier. All top-gates are set to -4V. This opens the barriers completely, because the measurement is done in the p-type region of the small bandgap SWNT. Fabry-Perot interference is observed over a wide gate range. The inset shows a zoom-in of the region in the white rectangle. One can clearly identify destructive interference at ~ ±0.6 meV (see arrows). The high conductance (up to G = 3.1 $e^2$/h) shows that the Pd contacts have a transmission close to T = 1.

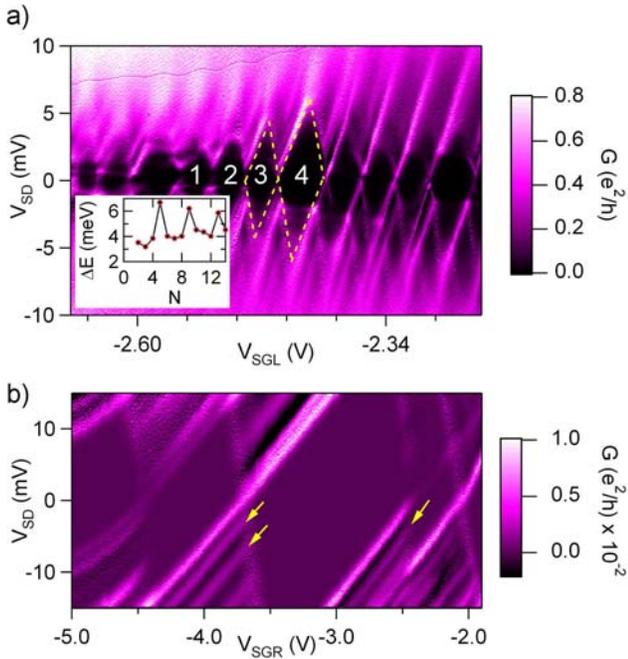

**Figure 3.** Stability diagrams of the left and right quantum dot: **a)** The left dot is formed with top-gate values $TG_L = TG_M = 4V$, $TG_R = -4V$. Four fold shell filling and excited states are clearly visible in the numeric dI/dV. In the inset, the addition energy of the $N^{th}$ electron is plotted for three shells. The yellow dotted lines are a guide to the eye indicating the shape of one small and one large diamond and the electron number in a particular shell is indicated. **b)** The right dot is formed with top-gate values $TG_L = -4 V$, $TG_M = 2V$, and $TG_R = 4V$. No obvious shell-filling is observed in this case. However, excited states are clearly visible. Those that belong to the level splitting $\Delta$ are indicated by arrows. Regions of negative differential conductance are also observed (black lines).

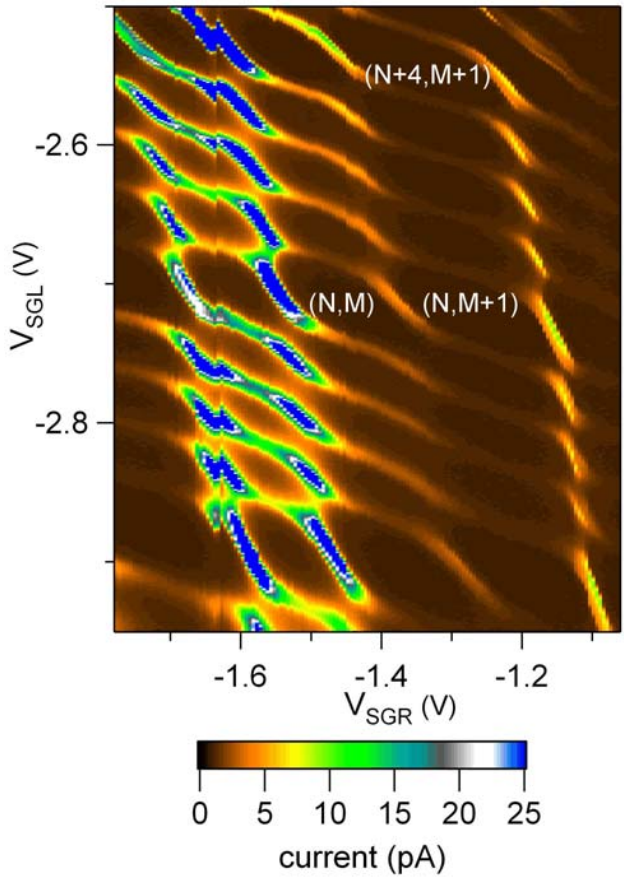

**Figure 4.** A double dot in the strongly coupled regime is formed with $TG_L = TG_M = 4V$, and $TG_R = 1.5V$ ($V_{SD} = 1mV$). The absolute values for (N,M) of the (left, right) quantum dot are unknown. However, the four-fold shell filling of the left dot is clearly visible in this honeycomb pattern, from which we can identify filled shells when (N, N+4, etc.).

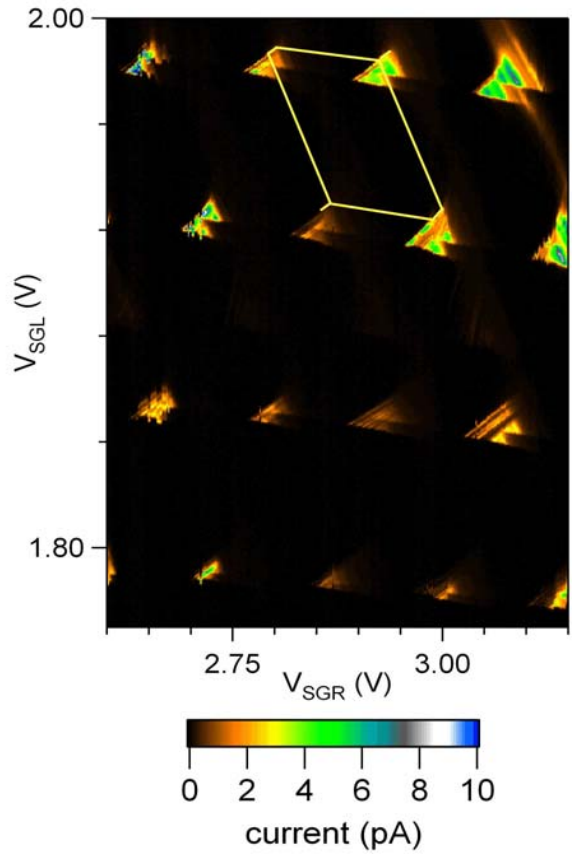

**Figure 5.** Characteristic "honeycomb" structure of the current for the double dot in the weakly coupled regime ($TG_L = TG_R = 0$ V, $TG_M = 200$ mV). The triple points with excited states are visible at the applied high bias of $V_{SD} = 5$ mV. From the size of the triangles and the hexagons, all capacitances that characterize the double dot can be calculated.

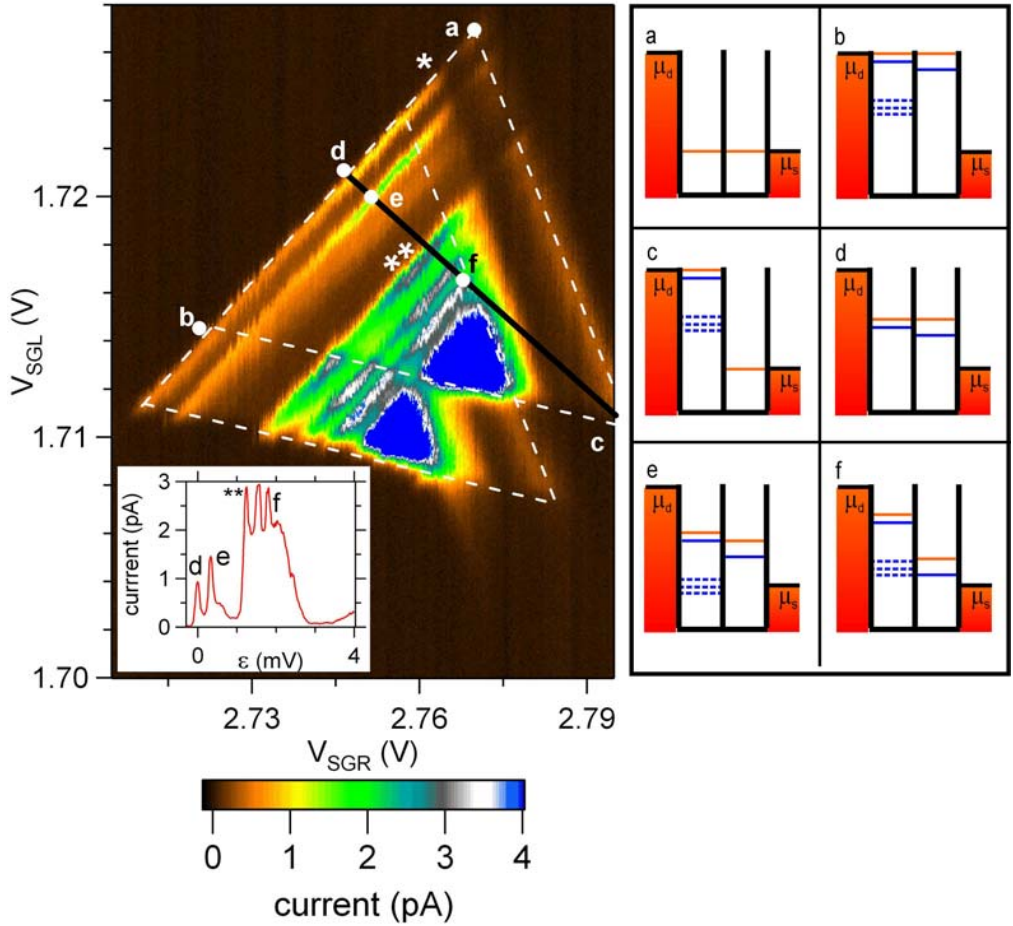

**Figure 6.** Current versus side gate voltages for a typical example for a triple point at high bias ($V_{SD}$ = 4 mV) in the p-doped region (hole transport). The ground state tunneling is weaker then tunneling through excited states. Lines parallel to the base (point **a** to point **b**) of a triangle belong to tunneling through excited states of the left dot, while the right dot excited states show up parallel to the upper-right side of the triangle (the first one enters the bias window at point (*)). The inset shows a line cut from the center of the base of the upper triangle to the triangle tip (black line from point **d** to point **c**), i.e., current as a function of the detuning ε between levels. The level schemes of the double dot corresponding to the points **a** to **f** depicted in the triangle are shown on the right side of the figure. Orange lines represent the ground states, blue lines hole-excited states (the measurement is done in the p-doped region). The dashed lines belong to the next excited states corresponding to (**).

(1) Loss, D.; DiVincenzo, D. P. *Phys. Rev. A* **1998**, 57, 120.

Kouwenhoven, L. P. *Science* **1998**, 282, 932.

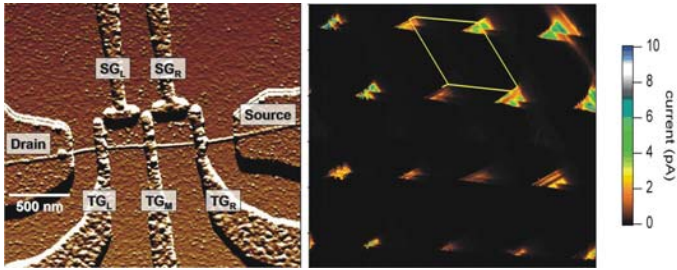